\documentclass[preprint,showpacs,aps]{revtex4}
\begin{document}
\tolerance=5000
\def\be{\begin{equation}}
\def\ee{\end{equation}}
\def\bea{\begin{eqnarray}}
\def\eea{\end{eqnarray}}

\title{Gravitational fields as generalized string models}

\author{Francisco J. Hern\'andez, Francisco Nettel and Hernando Quevedo}
\email{fmoreno, fnettel, quevedo@nucleares.unam.mx} 
\affiliation{   Instituto de Ciencias Nucleares,
     Universidad Nacional Aut\'onoma de M\'exico \\
     P.O. Box 70-543, 04510 M\'exico D.F., M\'exico}

\date{\today}

\begin{abstract}
We show that Einstein's main equations for 
stationary axisymmetric fields in vacuum are equivalent to  
the motion equations for bosonic strings moving on a special 
nonflat background. 
This new representation is based on the analysis 
of generalized harmonic maps in which the metric 
of the target space explicitly depends on the parametrization of
the base space. 
It is shown that this representation is valid 
for any gravitational field which possesses 
two commuting Killing vector fields. We introduce the
concept of dimensional extension which allows us to consider
this type of gravitational fields as strings embedded in D-dimensional
nonflat backgrounds, even in the limiting case where the Killing vector
fields are hypersurface orthogonal.

\end{abstract}
\maketitle
\section{Introduction}
\label{sec:int}
In general relativity,
stationary axisymmetric solutions of Einstein's equations 
play a crucial role for the description of the gravitational
field of astrophysical objects. In particular, the black hole
solutions and their generalizations that include Maxwell fields
are contained within this class. 

This type of exact solutions
has been the subject of intensive research during the past few
decades. In particular, the number of known exact solutions 
drastically increased after Ernst \cite{ernst} discovered an elegant 
representation of the field equations that made it possible  
to search for their symmetries. These studies led finally to   
the development of solution generating techniques \cite{solutions}
which allow us to find new solutions,
 starting from a given seed solution.
 In particular, solutions with an arbitrary number of 
multipole moments for the mass and angular momentum were
derived in \cite{ours} and used to describe the gravitational field
of rotating axially symmetric distributions of mass. 

The relationship between the Ernst representation and the 
Lagrangian for nonlinear sigma models \cite{misner,maison} was a 
further result that later allowed to formally develop the isomodromic
quantization procedure, based on the total integrability and
separation of variables of the underlying sigma model \cite{nicolai}.   
The analogy between sigma models and vacuum gravitational fields 
were further studied at the level of the action in \cite{cnq01}, where
it was shown that the Einstein-Hilbert Lagrangian for gravitational
fields with two commuting Killing vector fields can be reduced to 
the canonical Lagrangian of an $SL(2,R)/SO(2)$ nonlinear sigma model. 
Since string theories are based upon linear and nonlinear sigma models, 
some interesting relationships appear between 4-dimensional black hole solutions
and string solutions. In particular, in \cite{nishino1} it was shown that
any black hole solution can be at the same time the background of $N=2$ 
superstrings. Moreover, Tomimatsu-Sato generalizations of black holes can be
interpreted as closed string-like circular mass distributions as shown in 
\cite{nishino2}. The particular case of a Kerr--like solution of axidilaton 
gravity has been shown to possess a ring singularity in whose vicinity the 
solution can be interpreted as representing the field around a fundamental 
heterotic string \cite{burinskii}. 

In this work we derive an additional representation 
for stationary axisymmetric vacuum metrics. The 
fact that the main field equations for this field 
can  be derived from a Lagrangian 
with only two gravitational variables, depending on only two spacetime 
coordinates, is what we use as a starting point in order to develop
a formalism which allows us to interpret this special class of gravitational 
fields as generalized bosonic string models. The formalism is based
upon the definition of generalized harmonic maps which are characterized
by a new explicit connection between the metrics of the base space and the
target space. If the base space is $2-$dimensional, generalized harmonic 
maps can be interpreted as describing the motion of a bosonic string
on a nonflat background. 

Using the class of stationary axisymmetric vacuum fields as 
a prime example, we will show that any vacuum gravitational field with two
commuting Killing vector fields can be interpreted as a bosonic string 
``living" on a  curved background, whose metric explicitly depends on 
the parameters that are used to describe the string world-sheet. 
This interaction between the world-sheet metric and the background metric
manifests itself in the appearance of an additional term in the motion 
equations of the string and in 
a generalized conservation law for the
energy-momentum tensor of the string. These two new constituents of the
formalism allow us to identify the motion equations of the string 
with the main field equations of the gravitational field. This shows 
that Einstein's vacuum equations for this class of gravitational 
fields are equivalent to the motion equations of a generalized bosonic
string model. This analogy can be applied to the entire class of 
spacetimes  
with two commuting Killing vector fields, and we show it 
explicitly in the case of Einstein-Rosen gravitational waves and Gowdy 
cosmological models. Furthermore, particular sets of solutions contained
in this class of spacetimes are shown to satisfy boundary conditions
for open and closed strings.

This work is organized as follows. In Section \ref{sec:field} we 
introduce the notations for the stationary axisymmetric metric, review 
the $SL(2,R)/SO(2)$ representation of the main field equations, and 
show the incompatibility with the Polyakov action for bosonic strings.
In Section \ref{sec:examples} we present a generalization of harmonic 
maps which consists in considering metrics 
on the target space that explicitly depend on the coordinates 
of the base space. The mathematical properties of this new type
of harmonic maps are investigated in Appendix \ref{sec:sigma}.  
This generalization allows us to consider
a stationary axisymmetric gravitational field as described by 
a bosonic string moving on a nonflat background. Analogous results are
obtained for Einstein-Rosen waves and Gowdy cosmologies. 
Section \ref{sec:dimext}
is devoted to the discussion of a dimensional extension of the background space
which allows us to interpret a gravitational field of this class as a bosonic 
string moving on a nonflat space of arbitrary dimensions. 
Finally, Section \ref{sec:con} contains
the conclusions and suggestions for further research.

\section{Stationary axisymmetric gravitational fields}
\label{sec:field}
The first analysis of stationary axially symmetric gravitational fields
was carried out by Weyl \cite{weyl} in 1917, soon after the formulation of
general relativity. In particular, Weyl discovered that in the static limit
the main part of the vacuum field equations reduces to a single linear 
differential equation. The corresponding general solution can be written
in cylindrical coordinates as an infinite sum with arbitrary constant 
coefficients. A particular choice of the coefficients leads to the subset 
of asymptotically flat solutions which is the most interesting from 
a physical point of view. In this section we review the main properties 
of stationary axisymmetric gravitational fields. In particular, 
we show explicitly that the main field equations in vacuum can be 
represented as the equations of a nonlinear sigma model in which the
base space is the 4-dimensional spacetime and the target space is 
a 2-dimensional conformally Euclidean space.

\subsection{Line element and field equations} 
\label{sec:line}

Although there exist in the literature many suitable coordinate systems, 
stationary axisymmetric gravitational fields are usually described
in cylindric coordinates $(t,\rho,z,\varphi)$. Stationarity implies that
$t$ can be chosen as the time coordinate and the metric does not depend
on time, i.e. $\partial g_{ab}/\partial t =0$. Consequently, the corresponding 
timelike Killing vector has the components $\delta^a_t$. A second Killing
vector field is associated to the axial symmetry with respect to the axis 
$\rho=0$.  Then, choosing $\varphi$ as the azimuthal angle, the metric satisfies
the conditions $\partial g_{ab}/\partial \varphi =0$, and the components of 
the corresponding spacelike Killing vector are $\delta^a_\varphi$.

Using further the properties of stationarity and axial symmetry, together
with the vacuum field equations, for a general metric of the form 
$g_{ab}=g_{ab}(\rho,z)$, it is possible to show that the most general 
line element for this type of gravitational fields can be written
in the Weyl-Lewis-Papapetrou form as \cite{weyl,lewis,pap}
\be
ds^2 = f(dt-\omega d\varphi)^2 - 
f^{-1}\left[e^{2 k }(d\rho^2+dz^2) +\rho^2d \varphi^2\right] \ ,
\label{lel}
\ee
where $f$, $\omega$ and $k$ are functions of $\rho$ and $z$, only. 
After some rearrangements which include the introduction of a new function 
$\Omega=\Omega(\rho,z)$ by means of
\be
\rho \partial_\rho \Omega = f^2 \partial_z \omega \ , 
\qquad \rho \partial_z\Omega = - f^2   \partial_\rho \omega \ ,
\ee 
the vacuum field equations $R_{ab}=0$ can be
shown to be equivalent to the following set of partial differential equations
\be
\frac{1}{\rho}\partial_\rho(\rho\partial_\rho f)  + \partial_z^2 f + \frac{1}{f}[
(\partial_\rho \Omega) ^2  + (\partial_z \Omega) ^2 - (\partial_\rho f) ^ 2 -  
 (\partial_z f) ^2]
=0 \ ,
\label{main1}
\ee
\be
\frac{1}{\rho}\partial_\rho(\rho\partial_\rho \Omega)  + \partial_z^2 \Omega 
-  \frac{2}{f}\left(
\partial_\rho f\, \partial_\rho\Omega + \partial_z f \,\partial_z\Omega\right)
=0 \ ,
\label{main2}
\ee
\be
\partial_\rho k = \frac{\rho}{4f^2}\left[ (\partial_\rho f)^2+ (\partial_\rho \Omega)^2
 - (\partial_z f)^2  - (\partial_z \Omega)^2\right] \ ,
\label{krho}
\ee
\be
\partial_z k = \frac{\rho}{2f^2}\left( \partial_\rho f \ \partial_z f 
+ \partial_\rho \Omega \ \partial_z \Omega \right)\ .
\label{kz}
\ee
It is clear that the field equations for $k$ can be integrated by quadratures, 
once $f$ and $\Omega$ are known. For this reason, the equations 
(\ref{main1}) and (\ref{main2}) for $f$ and $\Omega$ are usually considered 
as the main field equations for stationary axisymmetric vacuum 
gravitational fields. In the following subsections we will focus on the analysis
of the main field equations, only. 

Let us consider the special case of static axisymmetric fields. This corresponds
to metrics which, apart from being axially symmetric and independent of the time 
coordinate, are invariant with respect to the transformation $\varphi \rightarrow 
-\varphi$ (i.e. rotations with respect to the axis of symmetry are not allowed). 
Consequently, the corresponding line element is given by (\ref{lel}) with 
$\omega=0$, and the field equations can be written as 
\be
\partial_\rho^2 \psi + \frac{1}{\rho}\partial_\rho \psi + \partial_z^2 \psi = 0 \ ,
\quad f=\exp(2\psi)\ ,\
\label{eqpsi}
\ee
\be
\partial_\rho k =  \rho\left[ (\partial_\rho \psi)^2 - (\partial_z \psi)^2\right]\ ,
\quad 
\partial_z k = 2 \rho  \partial_\rho \psi \ \partial_z \psi  \ . 
\label{eqkstatic}
\ee
We see that the main field equation (\ref{eqpsi}) corresponds to the linear 
Laplace equation for the metric function $\psi$. The general solution of 
Laplace's equation is known and, if we demand additionally asymptotic flatness, 
we obtain the Weyl solution which can be written as \cite{weyl,solutions} 
\be
\psi = \sum_{n=0}^\infty \frac{a_n}{(\rho^2+z^2)^\frac{n+1}{2}} P_n({\cos\theta}) \ ,
\qquad \cos\theta = \frac{z}{\sqrt{\rho^2+z^2}} \ ,
\label{weylsol}
\ee
where $a_n$ $(n=0,1,...)$ are arbitrary constants, and $P_n(\cos\theta)$ represents the Legendre
polynomials of degree $n$. 
The expression for the metric function $k$ 
can be calculated by quadratures by using the set of first order differential 
equations (\ref{eqkstatic}). Then 
\be
k = - \sum_{n,m=0}^\infty \frac{ a_na_m (n+1)(m+1)}{(n+m+2)(\rho^2+z^2)^\frac{n+m+2}{2} }
\left(P_nP_m - P_{n+1}P_{m+1} \right) \ .
\ee
Since this is the most general static, axisymmetric, asymptotically flat
vacuum solution, it must contain all known solution of this class. In particular,
one the most interesting special solutions which is Schwarzschild's spherically symmetric black hole spacetime must be contained in this class. To see this, we must choose
the constants $a_n$ in such a way that the infinite sum (\ref{weylsol}) converges to the
Schwarzschild solution in cylindric coordinates. But, or course, this representation 
is not the most appropriate to analyze the interesting physical properties of 
Schwarzchild's metric. 
 
\subsection{Representation as a nonlinear sigma model}
\label{sec:nlsm}
 
Consider two (pseudo)-Riemannian manifolds $(M,\gamma)$ and $(N, G)$ of dimension
$m$ and $n$, respectively. Let $M$ be coordinatized by $x^a$, and $N$ by $X^\mu$,
so that the metrics on $M$ and $N$ can be, in general, smooth functions of the 
corresponding coordinates, i.e., $\gamma=\gamma(x)$ and $G=G(X)$.  
A harmonic map is a smooth map $X: M \rightarrow N$, or in coordinates
$X: x \longmapsto X$ so that $X$ becomes a function of $x$, and the $X$'s  
satisfy the motion equations following from the action \cite{misner}
\be
S = \int d^m x \sqrt{|\gamma|}\ \gamma^{ab}(x)\ \partial_a X^\mu\  \partial_ b X^\nu \
G_{\mu\nu}(X)  \ ,
\label{acts}
\ee
which sometimes is called the ``energy" of the harmonic map $X$.
The straightforward variation of $S$ with respect to $X^\mu$ leads 
to the motion equations
\be
\frac{1}{\sqrt{|\gamma|}}\partial_b\left(\sqrt{|\gamma|}\gamma^{ab} \partial_a X ^\mu\right) 
+ \Gamma^\mu_{\ \nu\lambda} \ \gamma^{ab}\ \partial_a X^\nu \ \partial_b X^\lambda 
 = 0 \ ,
\label{moteq}
\ee
where $\Gamma^\mu_{\ \nu\lambda}$ are the Christoffel symbols associated to the 
metric $G_{\mu\nu}$ of the target space $N$. If $G_{\mu\nu}$ is a flat metric,
one can choose Cartesian-like coordinates such that $G_{\mu\nu}=\eta_{\mu\nu} =
{\rm diag}(\pm 1, ..., \pm 1)$, the motion equations become linear, and the
corresponding sigma model is linear. This is exactly the case of a bosonic 
string on a flat background in which the base space is the 2-dimensional string 
world-sheet. In this case the action (\ref{acts}) is usually referred to as 
the Polyakov action \cite{pol}.

Consider now the case in which the base space $M$ is a stationary axisymmetric 
spacetime. Then, $\gamma^{ab}$, $a,b =0,...,3$, can be chosen as
 the Weyl-Lewis-Papapetrou metric (\ref{lel}), i.e.
\be
\gamma_{ab} = \left( 
\begin{array}{cccc}
f & 0 & 0 & -f\omega \\
0 & - f^{-1} e ^{2k} & 0 & 0 \\
0 & 0 & - f^{-1} e ^{2k} & 0 \\
-f\omega & 0 & 0 & f \omega^2 - \rho^2 f^{-1} \\
\end{array}
\right)\ .
\ee 
Let the target space $N$ be 2-dimensional 
with metric $ G_{\mu\nu} = (1/2)f^{-2}\delta_{\mu\nu}$, $\mu,\nu=1,2$, and let 
the coordinates on $N$ be $X^\mu = (f,\Omega)$. Then, it is straightforward to 
show that the action  (\ref{acts}) becomes 
\be 
S = \int {\cal L}\, dt d\varphi d\rho dz \ , \qquad {\cal L} =
 \frac{\rho}{2f^2}\left[
(\partial_\rho f)^2 +  (\partial_z f)^2 +
 (\partial_\rho \Omega)^2 +  (\partial_z \Omega)^2\right]  \ ,
\label{lsig} 
\ee
and the corresponding motion equations (\ref{moteq})  
are identical to the main field equations (\ref{main1}) and (\ref{main2}). 

Notice that the field equations can also be obtained from (\ref{lsig}) 
by a direct variation with respect to $f$ and $\Omega$. This interesting 
result was obtained originally by Ernst \cite{ernst}, and is the starting
point of what today is known as the Ernst representation of the field 
equations.  

The above result shows that 
stationary axisymmetric gravitational fields can be described as a 
$(4\rightarrow 2)-$nonlinear harmonic map, where the base space is the 
spacetime of the gravitational field and the target space 
corresponds to a 2-dimensional conformally Euclidean space. A further analysis 
of the target space shows that it can be interpreted as the quotient space 
$SL(2,R)/SO(2)$ \cite{nicolai}, and the Lagrangian 
(\ref{lsig}) can be written explicitly \cite{cnq01} in terms of the generators 
of the Lie group $SL(2,R)$. Harmonic maps in which the target space is
a quotient space are usually known as nonlinear sigma models \cite{misner}.  

The form of the Lagrangian (\ref{lsig}) with two gravitational field variables, 
$f$ and $\Omega$, 
depending on two coordinates, $\rho$ and $z$, suggests a representation as a harmonic
map with a 2-dimensional base space. In string theory, this is 
an important fact that allows one to use the conformal invariance of the base 
space metric to find an adequate representation for the set of classical solutions. 
This, in turn, facilitates the application of the canonical quantization procedure. 
Unfortunately, this is not possible for the Lagrangian (\ref{lsig}).
Indeed, if we consider $\gamma^{ab}$ as a 2-dimensional metric that depends on 
the parameters $\rho$ and $z$, the diagonal form of the Lagrangian (\ref{lsig})
implies that $\sqrt{|\gamma|} \gamma^{ab} = \delta^{ab}$. Clearly, this choice is not 
compatible with the factor $\rho$ in front of the Lagrangian. Therefore, the 
reduced gravitational Lagrangian (\ref{lsig}) cannot be interpreted as corresponding
to a $(2\rightarrow n)$-harmonic map. 
Nevertheless, we will show 
in the next section that a modification of the definition of harmonic maps 
allows us to ``absorb" the unpleasant factor $\rho$ in the metric of the target space,  
and to use all the advantages of a 2-dimensional base space. 

Notice that the representation of stationary fields as a nonlinear sigma model becomes
degenerate in the limiting case of static fields. Indeed, 
the underlying geometric structure of the 
$SL(2,R)/SO(2)$ nonlinear sigma models requires that the target space be 2-dimensional, 
a condition which is not satisfied by static fields. We will see below that by using 
a dimensional extension of generalized sigma models, it will be possible to treat 
the special static case, without affecting the underlying geometric structure. 

The analysis performed in this section for stationary axisymmetric fields can be 
generalized to include any gravitational field containing two commuting 
Killing vector fields
\cite{solutions}. This is due to the fact that for this class of gravitational 
fields it is always possible to find the corresponding Ernst representation
in which the Lagrangian contains only two gravitational variables which depend
on only two spacetime coordinates.

\section{Gravitational fields as generalized harmonic maps}
\label{sec:examples}

A $(m\rightarrow n)-$generalized harmonic map is defined as a smooth map 
$X: M\rightarrow N$, satisfying the Euler-Lagrange equations 
\be
\frac{1}{\sqrt{|\gamma|}}\partial_b\left(\sqrt{|\gamma|}\gamma^{ab}
  \partial_a X^\mu \right) + \Gamma^\mu_{\ \nu\lambda } \, \gamma^{ab} \,
  \partial_a X^\nu  \partial_b X^\lambda + G^{\mu\lambda} \gamma^{ab} \, 
  \partial_a X^\nu \, \partial_b G_{\lambda\nu} 
 = 0 \ ,
\label{gengeo0}
\ee
which follow from the variation with respect to the fields $X^\mu$ 
of the generalized action  $ S = \int {\cal L} d^m x $ 
with the Lagrangian 
\be
{\cal L} = \sqrt{|\gamma|}\, \gamma^{ab}(x)\, \partial_a \, X^\mu
\partial_ b X^\nu  G_{\mu\nu}(X,x) \ .
\label{genlag0}
\ee
Here $(M,\gamma)$ and $(N,G)$ 
are (pseudo-)Riemannian manifolds of dimension $m$ and $n$, and
coordinates $x^a$ and $X^\mu$, respectively. Moreover, it is assumed
that $\gamma=\gamma(x)$ and $G=G(X,x)$, i.e. the target metric 
depends explicitly on the coordinates of the base space. This additional 
dependence is the result of the ``interaction" between the base space $M$
and the target space $N$, and leads to an extra term in the motion equations, 
as can be seen in (\ref{gengeo0}). In Appendix \ref{sec:sigma} we establish 
the main properties of generalized harmonic maps  which will be applied in 
concrete cases of gravitational fields in this section. First, we will 
analyze in detail the case of stationary axisymmetric fields and then we will
show that these results can be generalized to include other spacetimes with 
two commuting Killing vector fields, namely, the spacetimes of Einstein--Rosen 
gravitational waves and Gowdy cosmologies.

\subsection{Stationary axisymmetric spacetimes}
\label{sec:example}

In Section \ref{sec:field} we described stationary, axially symmetric, 
gravitational fields as a ($4 \to 2)-$nonlinear sigma model. There it was 
pointed out the convenience of having a 2-dimensional base space 
in analogy with string theory. Now we will show that this can be done 
by using the generalized harmonic maps defined above.

Consider a $(2 \to 2)-$generalized harmonic map. 
Let $x^a=(\rho,z)$ be the coordinates on the base space $M$,
and $X^\mu=(f,\Omega)$ the coordinates on the target space $N$.
In the base space we choose a flat metric and  in the target space 
a conformally flat metric, i.e. 
 \begin{equation} 
 \label{explmet}
\gamma_{ab} = \delta_{ab} \qquad \text{and} \qquad G_{\mu\nu}=\frac{\rho}{2f^2}\delta_{\mu\nu} \qquad (a,b=1,2; \ \mu,\nu=1,2).
\end{equation} 
A straightforward computation shows that the generalized Lagrangian (\ref{genlag0}) 
coincides with the Lagrangian (\ref{lsig}) for stationary axisymetric fields, and 
that the equations of motion (\ref{gengeo0}) generate the main field equations
(\ref{main1}) and (\ref{main2}).

For the sake of completeness we calculate the components of the energy-momentum 
tensor $ T _{ab}=\delta {\cal L}/\delta \gamma^{ab}$ (cf. Appendix \ref{sec:sigma}). 
Then
\begin{equation}   \label{emtaxi2}
T_{\rho\rho} = -T_{zz} =  \frac{\rho}{4f^2}\left[ (\partial_\rho f)^2 + (\partial_\rho \Omega)^2 - (\partial_z f)^2  - (\partial_z \Omega)^2 \right],
\end{equation}
\begin{equation}  
 \label{emtaxi3}
T_{\rho z} = \frac{\rho}{2f^2}\left( \partial_\rho f \,\partial_z f + \partial_\rho \Omega \,\,\partial_z \Omega \right).
\end{equation}
This tensor is traceless due to the fact that the base space is 2-dimensional. 
It satisfies the generalized conservation law (\ref{claw2})
on-shell:
\be 
\frac{d T_{\rho\rho}}{d \rho} + \frac{d T_{\rho z}}{d z} +\frac{1}{2}
\frac{\partial {\cal L}}{\partial \rho} = 0 \ ,
\label{claw3}
\ee
\be
\frac{d T_{\rho z}}{d \rho} - \frac{d T_{\rho \rho}}{d z} = 0 \ . 
\label{claw4}
\ee
Incidentally, the last equation coincides with the integrability condition for the metric
function $k$, which is identically satisfied by virtue of the main field equations. 
In fact, as can be seen from Eqs.(\ref{krho},\ref{kz}) and 
(\ref{emtaxi2},\ref{emtaxi3}), the components of the energy-momentum tensor 
satisfy the relationships 
$T_{\rho\rho} = \partial_\rho k$ and $ T_{\rho z} = \partial_z k$, so that 
the conservation law (\ref{claw4}) becomes an identity. Although we have eliminated
from the starting Lagrangian (\ref{lsig}) the variable $k$ by applying a Legendre 
transformation on the Einstein-Hilbert Lagrangian (see \cite{cnq01} for details)
for this type of gravitational fields, the formalism of generalized harmonic maps
seems to retain the information about $k$ at the level of the generalized 
conservation law.

The above results show that stationary axisymmetric spacetimes
can be represented as a $(2\to 2)-$generalized harmonic map with metrics given as
in (\ref{explmet}). 
It is also possible to interpret the generalized harmonic map given above 
as a generalized string model. Although the metric of the base space $M$ is Euclidean,
we can apply a Wick rotation $\tau=i\rho$ to obtain a Minkowski-like structure on $M$. 
Then, $M$ represents the world-sheet of a bosonic string in which $\tau$ is measures 
the time and $z$ is the parameter along the string. The string is ``embedded" in the
target space $N$ whose metric is conformally flat and explicitly depends on the 
time parameter $\tau$. We will see in the next section that this embedding becomes 
more plausible when the target space is subject to a dimensional extension.
In the present example, it is necessary to apply a Wick rotation in order to 
interpret the base space as a string world-sheet. This is due to the fact that 
both coordinates $\rho$ and $z$ are spatial coordinates. However, this can be
avoided by considering other classes of gravitational fields with timelike 
Killing vector fields; examples will be given below. 

The most studied solutions belonging to the class of stationary axisymmetric 
fields are the asymptotically flat solutions. Asymptotic flatness imposes conditions
on the metric functions which in the cylindrical coordinates used here can be 
formulated in the form
\be
\lim_{x^a\rightarrow\infty} f = 1 + O\left(\frac{1}{x^a}\right) \ ,\quad
\lim_{x^a\rightarrow\infty} \omega =  c_1 + O\left(\frac{1}{x^a}\right) \ , \quad
\lim_{x^a\rightarrow\infty} \Omega = O\left(\frac{1}{x^a}\right)
\ee
where $c_1$ is an arbitrary real constant which can be set to zero by appropriately 
choosing the angular coordinate $\varphi$. If we choose the domain of the spatial coordinates
as $\rho \in [0,\infty)$ and $z\in (-\infty, +\infty)$, from the asymptotic flatness 
conditions it follows that the coordinates of the target space $N$ satisfy the boundary 
conditions
\be
\dot X^\mu (\rho, -\infty) = 0 = \dot X^\mu (\rho, \infty) \ , \quad
  {X^\prime} ^\mu (\rho, -\infty) = 0 = {X^\prime}^\mu (\rho, \infty) \ 
\ee
where the dot stands for a derivative with respect to $\rho$ and the prime represents
derivation with respect to $z$. These relationships are known in string theory 
\cite{pol} as the Dirichlet and Neumann boundary conditions for open strings, 
respectively, with 
the extreme points situated at infinity. We thus conclude that if we assume $\rho$ as 
a ``time" parameter for stationary axisymmetric gravitational fields, an asymptotically 
flat solution corresponds to an open string  with endpoints attached to $D-$branes  
situated at plus and minus infinity in the $z-$direction. 


\subsection{Einstein--Rosen gravitational waves}
\label{sec:eiro}

Consider the line element for Einstein-Rosen gravitational waves \cite{solutions}
\be
ds^2 = e^{2(\gamma -\psi)} dt^2 - e^{-2\psi}(e^{2\gamma} d\rho^2 + \rho^2 d\varphi^2)
-e^{2\psi}(dz+\omega d\varphi)^2
\ee
where $\psi$, $\omega$ and $\gamma$ are functions of $t$ and $\rho$. These spacetimes
are characterized by the existence of two spacelike, commuting Killing vector fields 
$\xi^a_I = \delta^a_\varphi$ and   $\xi^a_{II} = \delta^a_z$. They describe the 
field of gravitational waves that propagate inward in vacuum,
implode on the axis of symmetry situated at $\rho=0$, and finally propagate outward to 
spatial infinity. The special case in which the Killing vectors are hypersurface orthogonal
corresponds to linearly polarized gravitational waves with $\omega =0$.

The reduced Einstein-Hilbert Lagrangian is obtained neglecting all the 
terms which can be represented as surface terms. The final result  
can be written as  \cite{cnq01}
\be
{\mathcal L}_{ER} =  2 \rho [(\partial_t\psi)^2 - (\partial_\rho\psi)^2] +\frac{1}{2}
\rho e^{-4\psi} [(\partial_t\Omega)^2 - (\partial_\rho\Omega)^2] \ ,
\label{lageiro}
\ee
where the function $\Omega$ is defined by $\rho \Omega_t = e^{4\psi} \omega_\rho$
and $\rho \Omega_\rho =  e^{4\psi} \omega_t$. The function $\gamma$ has been eliminated
by means of a Legendre transformation. Comparing the particular Lagrangian 
(\ref{lageiro}) with the general Lagrangian (\ref{genlag0}), it is easy to establish 
that it corresponds to a $(2\rightarrow 2)-$generalized harmonic map with a Minkowski-like
base space $(M,\gamma(x))$, i. e.,
\be
x^1= t\ , \ x^2 = \rho\ ,  \
\gamma_{ab} = {\rm diag}(1,-1) \ ,
\ee
and a curved target space $(N,G(X,x))$ with 
\be
 X^1 = \psi \ , \ X^2 = \Omega\ ,
G_{\mu\nu} = {\rm diag}\left[ 2\rho, (\rho/2)e^{-4\psi}\right]  \ .
\ee
The field equations (\ref{gengeo0}) for this particular generalized harmonic 
map can be written as  
\be
\partial_\rho^2\psi + \frac{1}{\rho}\partial_\rho \psi -\partial_t^2 \psi
+\frac{1}{2}e^{-4\psi}[(\partial_\rho\Omega)^2 - (\partial_t\Omega)^2] =0 \ ,
\ee
\be
\partial_\rho^2\Omega + \frac{1}{\rho}\partial_\rho \Omega -\partial_t^2 \Omega
+4[\partial_t \Omega\ \partial_t\psi - \partial_\rho \Omega\ \partial_\rho\psi] = 0 \ ,
\ee 
and coincide with the main Einstein's field equations in empty space for this kind
of gravitational waves.

As for the energy-momentum tensor associated with the string metric $\gamma_{ab}$, 
the components read
\be
T_{tt} = T_{\rho\rho}=  \rho [(\partial_t\psi)^2 +(\partial_\rho\psi)^2] +\frac{1}{4}
\rho e^{-4\psi} [(\partial_t\Omega)^2+ (\partial_\rho\Omega)^2] \ ,
\ee
\be
T_{t\rho} = 2 \rho \partial_t \psi\ \partial_\rho\psi + \frac{1}{2}\rho e^{-4\psi}
\partial_t \Omega\  \partial_\rho\Omega \ .
\ee
Finally, it can be shown that 
$\partial_\rho\gamma = T_{tt}$ and $\partial_t\gamma = T_{t\rho}$ 
so that the integrability condition for the function $\gamma$ corresponds to the 
generalized conservation law (\ref{claw2}).  

The above results show that Einstein-Rosen gravitational waves 
can be interpreted as a particular generalized harmonic map and that particular solutions
of the field equations correspond to a string spatially situated along the coordinate 
$\rho$ and moving along the time coordinate $t$. The string propagates on a 
2-dimensional nonflat background with metric $G$.  As for the boundary conditions
of this type of strings, if we choose a particular wave solution with a regular curvature 
behavior everywhere in spacetime, except at the wave front, the metric functions
$\psi$, $\omega$ and $\Omega$ must satisfy certain relationships (see, for
instance, \cite{patquev05}) which can be 
expressed as the Dirichlet and Neumann conditions for and open string 
in the form
\be
\dot X^\mu (t, 0) = 0 = \dot X^\mu (t, \infty) \ , \quad
  {X^\prime} ^\mu (t, 0 ) = 0 = {X^\prime}^\mu (t, \infty) \ . 
\ee
Here the dot stands for a derivative with respect to the time coordinate $t$ 
and the prime represents derivation with respect to the spatial coordinate $\rho$. 
The endpoints are situated on the axis of symmetry, $\rho =0$, and at infinity.
We see that an Einstein-Rosen gravitational wave can be interpreted as an open 
string attached to $D-$branes located on the axis and at infinity in the $\rho-$direction.
Since the wave propagates inwards and outwards in empty space, its singular front
reaches the endpoints at some moment, say at $t_0$ and at $t_\infty$, where the metric
and its curvature diverge so that the analogy with $D-$branes breaks down.

\subsection{Gowdy cosmological models}
\label{sec:gow}

Consider the Gowdy cosmological models whose line element in the unpolarized 
$T^3$ case can be written as \cite{gow}
\be
ds^2 = e^{-(\lambda+3\tau)/2} d\tau^2 - e^{-(\lambda - \tau)/2} d\theta^2
- e^{-\tau}[e^P(d\sigma+Q d\delta)^2 + e^{-P}d\delta^2] \ ,
\ee
where $P$, $Q$, and $\lambda$ are functions of $\tau$ and $\theta$ only.
The spacelike Killing vector fields are associated to the coordinates $\sigma$ and
$\delta$, i. e., $\xi^a_I = \delta^a_\sigma$ and $\xi^a_{II} = \delta^a_\delta$.  These
spacetimes are the simplest, inhomogeneous, spatially closed cosmological models in 
vacuum. They are expected to describe the geometric behavior of cosmological 
inhomogeneities and are useful in the study of the geometric properties of initial 
cosmological singularities. The special case in which $Q=0$ is usually known as 
the polarized model and corresponds to the 
limiting case of hypersurface orthogonal Killing
vectors.

The reduced Einstein-Hilbert Lagrangian can be expressed as  
\be
{\mathcal L}_{Gow} = \frac{1}{2} \left\{(\partial_\tau P)^2 - e^{-2\tau} (\partial_\theta P)^2 ]
+e^{2P}[ (\partial_\tau Q)^2 - e^{-2\tau} (\partial_\theta Q)^2]\right\} \ ,
\label{laggow}
\ee
where a Legendre transformation has been used to eliminate the cyclic function $\lambda$
\cite{cnq01}. The corresponding field equations can be obtained by varying this Lagrangian 
density with respect to $P$ and $Q$ independently. As in the previous examples, 
to establish the relationship 
with generalized harmonic maps, we compare the particular Lagrangian 
(\ref{laggow}) with the general Lagrangian (\ref{genlag0}). It is then easy to identify
the coordinates and metric of the base space $M$ as
 \be
x^1= \tau\ , \ x^2 = \theta \ , \  
\gamma^{ab} = {\rm diag}(1,-e^{-2\tau}) \ ,
\label{bsgow}
\ee
and of the target space $N$ as
\be
X^1 = P \ , \ X^2 = Q\ , \
G_{\mu\nu} = \frac{1}{2} e^{-\tau}{\rm diag}(1,e^{2P}) \ . 
\ee
Moreover, the motion equations motion equations (\ref{gengeo0}) lead to the set  
\be
\partial_\tau^2 P - e^{-2\tau}\partial_\theta^2 P - e^{2P} 
[(\partial_\tau Q)^2 - e^{-2\tau}(\partial_\theta Q)^2] = 0\ ,
\ee
\be
\partial_\tau^2 Q - e^{-2\tau}\partial_\theta^2 Q + 2 
[\partial_\tau P \ \partial_\tau Q - e^{-2\tau}\partial_\theta P \ \partial_\theta Q] = 0\ ,
\ee
which are equivalent to the main Einstein field equations in empty space. 

Using Eq.(\ref{bsgow}), the base space $(M,\gamma(x))$ in this case can be shown to correspond to a 2-dimensional pseudo-Riemannian manifold of negative 
constant curvature, whereas
the target manifold $(N,G(X,x))$ is in general characterized by a non-constant 
curvature. This means that any Gowdy cosmological model is at the same
time a string with constant local curvature which propagates on a 2-dimensional 
curved background space.  

Finally, the components of
the energy-momentum tensor $T_{ab}$ (cf. Eq.(\ref{emt})) 
are 
\be
T_{\tau\tau}=e^{-2\tau}T_{\theta\theta} = \frac{1}{4}\left\{
(\partial_\tau P)^2 + e^{-2\tau}(\partial_\theta P)^2 + e^{2P}[
(\partial_\tau Q)^2 + e^{-2\tau}(\partial_\theta Q)^2]\right\}\ ,
\ee
\be
T_{\tau\theta} = \frac{1}{2} \left( \partial_\tau P\ \partial_\theta P 
+ e^{2P}\partial_\tau Q\ \partial_\theta Q \right)  \ .
\ee
The generalized conservation law (\ref{claw2}) 
is equivalent to the integrability condition 
for the function $\lambda$, since $\partial_\tau\lambda = 4 T_{\tau\tau}$ and
$\partial_\theta\lambda = 4 T_{\tau\theta}$. Moreover, the condition $T^a_{\ a}=0$ 
is identically satisfied due to the conformal invariance of the string metric.

We will now establish a relationship between a class of Gowdy cosmologies 
and the boundary conditions of the string. The most important class of 
Gowdy spacetimes are the so called asymptotically velocity term dominated
(AVTD) cosmologies which are expected to describe the initial cosmological 
singularity $(\tau\rightarrow\infty)$ 
from a geometrical point of view \cite{bkz}. It can be shown 
that AVTD cosmologies behave at $\tau\rightarrow\infty$ as \cite{smq} 
\be
P = \ln [ a( e^{-c\tau} + b^2 e^{c\tau})]\ , \qquad
Q = {b \over a(e^{-2c\tau} + b^2)} + d \ ,
\label{avtd}
\ee
where $a ,\ b, \ c$ and $d$ are arbitrary real functions of $\theta$. Since the
angular coordinate $\theta$ is defined in the interval $[0,2\pi]$, 
from the functional dependence of the string metric $\gamma_{ab}$ and 
from the arbitrariness of the functions entering the AVTD expressions for $P$ and $Q$ in 
(\ref{avtd}), it can be shown that in this case the boundary conditions for a closed
string \cite{pol} 
\be
\gamma_{ab}(\tau,0)=\gamma_{ab}(\tau,2\pi)\ ,\quad
X^\mu(\tau,0) = X^\mu (\tau, 2\pi) \ ,\quad
{X^\prime}^\mu(\tau,0) = {X^\prime}^\mu (\tau, 2\pi) \ ,
\ee
are satisfied. Here the prime denotes differentiation with respect to $\theta$. 
This results establishes that an AVTD Gowdy cosmology can be interpreted as 
a closed string with a constant curvature geometry propagating on a nonflat
background.

\section{Dimensional extension}
\label{sec:dimext}

In order to further analyze the analogy between gravitational fields and bosonic string 
models, we perform an arbitrary dimensional extension of the target space $N$, and
study the conditions under which this dimensional extension 
does not affect the field equations of the gravitational field.
Consider an $(m\rightarrow D)-$generalized harmonic map. As before we denote by 
$\{x^a\}$ the coordinates on $M$. Let $\{X^\mu, X^\alpha\}$ with $\mu=1,2$ and $\alpha=3,4,..., D$ be the coordinates on $N$. The metric structure 
on $M$ is again $\gamma = \gamma(x)$, 
whereas the metric on $N$ can in general depend on all coordinates of
$M$ and $N$, i.e. $G=G(X^\mu,X^\alpha,x^a)$. 
The general structure of the corresponding field equations is as given in 
(\ref{gengeo0}). They can be divided into one set of equations for $X^\mu$ and one set 
of equations for $X^\alpha$. According to the results of the last section, 
the class of gravitational fields under consideration can be represented as a 
$(2\rightarrow 2)-$generalized harmonic map so that we can assume that 
the main gravitational variables are contained in the coordinates $X^\mu$ of the
target space. Then, the gravitational sector of the target space
will be contained in the components $G_{\mu\nu}$ ($\mu,\nu=1,2$) of the metric, 
whereas the components $G_{\alpha\beta}$ $(\alpha,\beta=3,4,...,D)$ represent
the sector of the dimensional extension. 

Clearly, the set of differential 
equations for $X^\mu$ also contains the variables
$X^\alpha$ and its derivatives $\partial_a X^\alpha$. For the gravitational 
field equations to remain unaffected by this dimensional extension we demand 
the vanishing of all the terms containing $X^\alpha$ and its derivatives 
in the equations for $X^\mu$. It is easy to show that this can be achieved 
by imposing the conditions 
\be
G_{\mu\alpha}=0\ , \quad \frac{\partial G_{\mu\nu}}{\partial X^\alpha} =0 \ ,
\quad \frac{\partial G_{\alpha\beta}}{\partial X^\mu}=0 \ .
\label{cond}
\ee
That is to say that the gravitational sector must remain completely invariant 
under a dimensional extension, and the additional sector cannot depend on the
gravitational variables, i.e., $G_{\alpha\beta}= G_{\alpha\beta}(X^\gamma, x^a)$, 
$\gamma=3,4,...,D$.  
Furthermore, the variables $X^\alpha$ must satisfy the differential equations
\be
\frac{1}{\sqrt{|\gamma|}}\partial_b\left(\sqrt{|\gamma|}\gamma^{ab}
  \partial_a X^\alpha \right) + \Gamma^\alpha_{\ \beta\gamma } \, \gamma^{ab} \,
  \partial_a X^\beta  \partial_b X^\gamma + G^{\alpha\beta} \gamma^{ab} \, 
  \partial_a X^\gamma \, \partial_b G_{\beta\gamma} 
 = 0 \ .
\ee
This shows that any given $(2\rightarrow 2)-$generalized map can be extended, without
affecting the field equations, to a
$(2\rightarrow D)-$generalized harmonic map. 

It is worth mentioning that the fact 
that the target space $N$ 
becomes split in two separate parts implies that the 
energy-momentum tensor $T_{ab} = \delta {\cal L}/\delta \gamma^{ab}$ separates into 
one part belonging to the gravitational sector and a second one following from the
dimensional extension, i.e. $T_{ab} =  T_{ab}(X^\mu,x) +  T_{ab}(X^\alpha,x)$. 
The generalized conservation law as given in (\ref{claw2}) is 
satisfied by the sum of both parts.

Consider the example of stationary axisymmetric fields given the metrics
(\ref{explmet}). Taking into account the conditions (\ref{cond}), after a 
 dimensional extension  the metric of the target space becomes 
\begin{equation}
G = 
\left( \begin{array}{ccccc} 
\frac{\rho}{2f^2}      & 0 &  0  & \cdots &   0  \\
0      & \frac{\rho}{2f^2} &  0  & \cdots &   0  \\
0  &   0  & G_{33}(X^\alpha,x) & \cdots  &  G_{3D}(X^\alpha,x)  \\
.  &  .    & \cdots            & \cdots  & \cdots \\
0  & 0 & G_{D3}(X^\alpha,x) & \cdots    & G_{DD}(X^\alpha,x) 
\end{array}\right).
\end{equation}
Clearly, to avoid that this metric becomes degenerate we must demand that 
$\det(G_{\alpha\beta})\neq 0$, a condition that can be satisfied in view of
the arbitrariness of the components of the metric. With the extended metric,
the Lagrangian density gets an additional term
\begin{equation}
\mathcal{L} = \frac{\rho}{2f^2}\left[(\partial_\rho f)^2 +  (\partial_z f)^2 + (\partial_\rho \Omega)^2 +  (\partial_z \Omega)^2\right] +  
\left(\partial_\rho X^\alpha \partial_\rho X^\beta 
+ \partial_z X^\alpha \partial_z X^\beta\right)G_{\alpha\beta} \ ,
\end{equation}
which nevertheless does not affect the field equations for the gravitational 
variables $f$ and $\Omega$. 
On the other hand, 
the new fields must be solutions of the extra field equations
\begin{equation}
\left(\partial_{\rho}^2 +\partial_{z}^2\right)X^\alpha + \Gamma^{\alpha}_{\;\;\beta\gamma}
\left(\partial_\rho X^\beta \partial_\rho X^\gamma 
       + \partial_z X^\beta \partial_z X^\gamma \right) 
       + G^{\alpha\gamma}\left(\partial_\rho X^\beta \partial_\rho G_{\beta\gamma} 
       + \partial_z X^\beta \partial_z G_{\beta\gamma}\right) = 0 \ .
\end{equation}

An interesting special case of the dimensional extension is the one in which the extended
sector is Minkowskian, i.e. for the choice $G_{\alpha\beta}=\eta_{\alpha\beta}$ with 
additional fields $X^\alpha$ given as arbitrary harmonic functions. This choice opens 
the possibility of introducing a ``time" coordinate as one of the additional dimensions, 
an issue that could be helpful when dealing with the interpretation of gravitational
fields in this new representation. 

The dimensional extension finds an interesting application in the case of static 
axisymmetric gravitational fields. As mentioned in Section \ref{sec:line}, these
fields are obtained from the general stationary fields 
in the limiting case $\Omega =0$ (or equivalently, $\omega=0)$. 
If we consider the representation as an $SL(2,R)/SO(2)$ nonlinear sigma model 
or as a $(2 \to 2)-$generalized harmonic map, we see immediately that the
limit $\Omega=0$ is not allowed because the target space becomes 1-dimensional 
and the underlying metric is undefined. To avoid this degeneracy, we first apply 
a dimensional extension and only then calculate
the limiting case $\Omega=0$. In the most simple case of an extension with 
$G_{\alpha\beta}=\delta_{\alpha\beta}$, the resulting $(2\to 2)-$generalized
map is described by the metrics $
\gamma_{ab}=\delta_{ab}$ 
and 
\be
G = 
\left(\begin{array}{cc}
\frac{\rho}{2f^2}      & 0  \\
0      & 1
\end{array}\right)
\ee
where the additional dimension is coordinatized by an arbitrary harmonic function
which does not affect the field equations of the only remaining gravitational variable $f$. This scheme represents 
an alternative method for exploring static fields on  nondegenerate target spaces.
Clearly, this scheme can be applied  to the gravitational fields mentioned in the Appendix and, in general, to the case of gravitational fields possessing two hypersurface orthogonal Killing vector fields.

Our results show that a stationary axisymmetric field can be represented as  
a string ``living" in a $D$-dimensional target space $N$. 
The string world-sheet is parametrized by the coordinates $\rho$ and $z$. The 
gravitational sector of the target space depends explicitly on the metric functions
$f$ and $\Omega$ and on the parameter $\rho$ of the string world-sheet. The sector
corresponding to the dimensional extension can be chosen as a $(D-2)-$dimensional 
Minkowski spacetime with time parameter $\tau$. Then, the string world-sheet is
a 2-dimensional flat hypersurface which is ``frozen" along the time $\tau$.

\section{Conclusions}
\label{sec:con}

In this work, we introduced the concept of generalized harmonic maps which are
characterized by a new explicit interaction between the metric of the base space and 
the metric of the target space. This interaction is realized by means of an explicit 
dependence of the target space metric in terms of the coordinates of the base space. 
The action of the generalized harmonic map becomes directly influenced by the
existence of the additional interaction. 
As a result of this new dependence, an additional term appears in the differential 
equations that determine the harmonic map. Furthermore, a generalized 
conservation law is satisfied by the energy-momentum tensor obtained by varying
the action with respect to the metric of the base space. In the case of a 2-dimensional 
base space we interpret a generalized map as describing the behavior of a string
embedded in the target space. 

We showed that any vacuum gravitational field with two
commuting Killing vector fields accepts a representation as a $(2\to 2)-$generalized
harmonic map and, consequently, can be interpreted as a bosonic string 
``living" on a  curved background, whose metric explicitly depends on 
the parameters that are used to describe the string world-sheet. 
This result indicates that 
Einstein's vacuum equations for this class of gravitational 
fields are equivalent to the motion equations of a generalized bosonic
string model. The case of stationary axisymmetric vacuum fields was 
used throughout the work to illustrate the details of this new
representation. In this particular example we saw that the
base space is flat and the target space defines a conformally 
flat background. Moreover, in the case of 
Einstein-Rosen gravitational waves and a class of 
Gowdy cosmological models the reinterpretation in terms of 
generalized string models holds, with more 
general metrics for the
base space and the target space. It was shown that physical conditions
imposed on the behavior of the spacetime metrics correspond to boundary 
conditions on the string models. For instance, asymptotic flatness
in stationary axisymmetric spacetimes corresponds to  Dirichlet or Neumann
boundary
conditions for an open string with endpoints situated at infinity. A regular
Einstein-Rosen gravitational wave can be interpreted as an open string 
with endpoints localized at the symmetry axis and at infinity. Finally, 
the so called asymptotically velocity term dominated (AVTD) Gowdy cosmologies
are at the same time closed strings 
 with a constant curvature geometry,  propagating on a nonflat
background. We expect that this analogy between physical conditions of 
the spacetime metrics and boundary conditions of the string models holds 
in more general cases.

Our approach allows a dimensional extension in which the class of 
gravitational fields with two commuting Killing vectors can be 
represented as $(2\to D)-$generalized harmonic maps. In particular,
we used this extension to show that it is possible to investigate
the limiting case of static gravitational fields as a generalized
map, avoiding the problem of the degeneracy of the target space.

It would be interesting to investigate the possibility of using
the present representation in the context of canonical quantization. 
In fact, one important result of string theory is that when one
quantizes a string on a flat background, one obtains an infinite
tower of massive states which are partially identified with 
elementary particles. In our case, however, gravitational fields
are represented by strings moving on nonflat backgrounds. 
Furthermore, one of the main reasons why the canonical quantization 
of the bosonic string 
on curved backgrounds presents serious difficulties is because exact 
solutions of the corresponding field equations are very difficult to be found
\cite{san}. 
Nevertheless, for the gravitational fields under consideration, 
this problem has already been solved. In fact, the special case of static solutions 
can be solved in general, as we mentioned in Section \ref{sec:line}. 
Solution 
generating techniques \cite{solutions} can be used to find the general stationary solution, 
for instance, in terms of multipole moments \cite{ours}. We believe that 
this advantage can be used to formulate quantization schemes for this special class of 
gravitational fields.

\begin{acknowledgments}
This work was supported in part by CONACyT, grant 48601-F. 
\end{acknowledgments}

\appendix  
\section{Generalized harmonic maps}
\label{sec:sigma}

Consider two (pseudo-)Riemannian manifolds $(M,\gamma)$ and $(N,G)$ 
of dimension $m$ and $n$, respectively.
Let $x^a$ and $X^\mu$ be coordinates on $M$ and $N$, respectively. 
This coordinatization implies that in general the 
metrics $\gamma$ and $G$ become functions of the corresponding 
coordinates. Let us assume that not only $\gamma$ 
but also $G$ can explicitly depend on the coordinates $x^a$, i.e.
let $\gamma=\gamma(x)$ and $G=G(X,x)$. This simple assumption 
is the main aspect of our generalization which, as we will see, lead
to new and nontrivial results. 

A smooth map $X: M\rightarrow N$ will be called an 
$(m\rightarrow n)-$generalized harmonic
map if it satisfies the Euler-Lagrange equations 
\be
\frac{1}{\sqrt{|\gamma|}}\partial_b\left(\sqrt{|\gamma|}\gamma^{ab}
  \partial_a X^\mu \right) + \Gamma^\mu_{\ \nu\lambda } \, \gamma^{ab} \,
  \partial_a X^\nu  \partial_b X^\lambda + G^{\mu\lambda} \gamma^{ab} \, 
  \partial_a X^\nu \, \partial_b G_{\lambda\nu} 
 = 0 \ ,
\label{gengeo}
\ee
which follow from the variation of the generalized action  
\be  
\label{genact}
S = \int d^m x \sqrt{|\gamma|}\, \gamma^{ab}(x)\, \partial_a \, X^\mu
\partial_ b X^\nu  G_{\mu\nu}(X,x)   \ ,
\ee
with respect to the fields $X^\mu$. Here the Christoffel symbols, 
determined by the metric $G_{\mu\nu}$, are calculated in the standard 
manner, without considering the explicit dependence on $x$.  
Notice that the new ingredient in this generalized definition of harmonic maps,
i.e.,  the
term $G_{\mu\nu}(X,x)$ in the Lagrangian density implies that we are taking 
into account the ``interaction" between the base space $M$ and the target
space $N$. This interaction leads to an extra term in the motion equations, 
as can be seen in (\ref{gengeo}).
It turns out that this interaction is the result of the effective 
presence of the gravitational field. 

Notice that the limiting case of generalized linear harmonic maps is
much more complicated than in the standard case. Indeed, for the motion 
equations (\ref{gengeo}) to become linear it is necessary that the
conditions 
\be 
\gamma^{ab} ( 
\Gamma^\mu_{\ \nu\lambda } \, \,
    \partial_b X^\lambda + G^{\mu\lambda} \, 
  \partial_b G_{\lambda\nu} )\partial_a X^\nu
 = 0 \ ,
\label{lincon}
\ee
be satisfied. One could search for a solution in which 
each term vanishes  separately. 
The choice of a (pseudo-)Euclidean 
 target metric $G_{\mu\nu}=\eta_{\mu\nu}$, which would
 imply $\Gamma^\mu _{\ \nu\lambda}=0$, 
is not allowed, because it would contradict the assumption  
$\partial_b G_{\mu\nu} \neq 0$. Nevertheless, a flat background metric
in curvilinear coordinates could be chosen such that  the 
assumption $G^{\mu\lambda}\partial_b G_{\mu\nu} =  0$ is fulfilled, but in this case
$\Gamma^\mu _{\ \nu\lambda}\neq 0$ and (\ref{lincon}) cannot be satisfied. 
In the general case of a curved 
target metric, conditions (\ref{lincon}) represent a system of $m$
 first order nonlinear partial differential equations for $G_{\mu\nu}$. Solutions
to this system would represent linear generalized harmonic maps. The
complexity of this system suggests that this special type of maps
is not common.   

\subsection{Symmetries of the action}
\label{sec:sym}

Let us consider the symmetries of the generalized action (\ref{genact}), i.e. 
transformations involving the ``variables" $x,\ X,\ \gamma$ and $G$ 
such that $S$ remains invariant.  

The first obvious symmetry follows from the application of 
diffeomorphisms of the target space, 
\begin{equation}
X^\mu \to X'^\mu = X'^\mu(X),
\label{inftra}
\end{equation}
which leave invariant the metric structure of the base space $\gamma=\gamma(x)$, 
but they affect the metric $G$ of the target space and the partial derivatives 
of the fields $X^\mu$,
\begin{equation}
G'_{\mu\nu} = \frac{\partial X^\alpha}{\partial X'^\mu}\frac{\partial X^\beta}{\partial X'^\mu}G_{\alpha\beta}, \qquad \text{and} \qquad \partial_a X'^\mu = \frac{\partial X'^\mu}{\partial X^\beta}\partial_a X^\beta.
\end{equation}
Then, it follows that the form of the Lagrangian density of the action (\ref{genact}) is left unchanged.

The diffeomorphism invariance or reparametrization of the base space $x^a \to x'^a = x'^a(x)$ requires more attention because of the explicit dependence of the metric of the target space on the coordinates of the base space, $G_{\mu\nu} = G_{\mu\nu}(X, x)$.
The volume element $\sqrt{|\gamma|}d^m x$ in (\ref{genact}) is by definition an invariant.
Let us introduce the notation 
\be
 h_{ab} = \partial_a \, X^\mu \partial_ b X^\nu G_{\mu\nu}(X,x) \ ,
\ee
so that the integrand of (\ref{genact}) can be written as $\gamma^{ab}(x) h_{ab} (X,x)$.
By construction, we know that the contravariant form of the metric of the base space $\gamma^{ab}$ transforms as a $(2,0)$ rank tensor. Then, the expression 
$\gamma^{ab}h_{ab}$ will transform as a scalar only if $h_{ab}$ transforms as a $(0,2)$ 
tensor. In other words, the invariance of the generalized action (\ref{genact}) is
fulfilled if $h_{ab}(X,x)$ is a $(0,2)$ tensor which corresponds to the metric 
induced on the target space $N$ by means of the map $X: M \to N$. This is equivalent
to say that $h=X^*(G)$, where $X^*$ is the pullback associated to the map 
$X$. Now we will 
show that in fact $h_{ab}$ transforms as the components of an induced metric. Let us
recall that the transformation law
\begin{equation}  
\label{trgamma}
\gamma'^{ab} = \frac{\partial x'^a}{\partial x^c}\frac{\partial x'^b}{\partial x^d}\gamma^{cd} \ ,
\end{equation}
implies for an infinitesimal diffeomorphism  
\be
x^a \to x'^a = x^a + \epsilon \xi^a(x) \ ,
\ee
where $\epsilon$ is an infinitesimal parameter, that 
\begin{equation}
\gamma^{ab} \to \gamma'^{ab} = \gamma^{ab} + \epsilon ( \partial_c \gamma^{ab} \xi^c
-\gamma^{cb}\partial_c\xi^a - \gamma^{ac}\partial_c\xi^b )
= \gamma^{ab} + \epsilon \mathsterling_\xi \gamma^{ab} \ .
\label{lie}
\end{equation}
Here $\mathsterling_\xi$ is the Lie derivative with respect to the vector field $\xi^a$ tangent to the integral curves of the diffeomorphism. Vice versa, if the expression 
$\gamma^{ab}$ transforms under an infinitesimal diffeomorphism  as in (\ref{lie}), 
it can be shown that the corresponding finite diffeomorphism satisfies 
the transformation law (\ref{trgamma}). For the components of $h_{ab}$ consider 
the geometric object
\be
h = h_{ab} dx^a dx^b = G_{\mu\nu}(X,x)  \partial_a \, X^\mu \partial_ b X^\nu 
dx^a dx^b \ .
\label{hobj}
\ee
It is straightforward to show that under an infinitesimal diffeomorphism 
of the form (\ref{inftra}), the expressions entering this object transform as
\be
X^\mu(x) \to  X^\mu(x') = X^\mu(x) + \epsilon \partial_c X^\mu(x) \xi^c \ ,
\ee
\be
G_{\mu\nu}(X(x),x) \to G_{\mu\nu}(X(x'),x') = G_{\mu\nu}(X(x),x) + \epsilon \left( \partial_\lambda G_{\mu\nu} \partial_c X^\lambda \xi^c + \partial_c G_{\mu\nu} \xi^c \right).
\ee
Applying now the infinitesimal diffeomorphism to $h$ as given in (\ref{hobj}), 
and considering only terms up to the first order in $\epsilon$, 
after some algebraic manipulations we obtain 
\bea
h' =  \{&&  G_{\mu\nu}\partial_a X^\mu \partial_b X^\nu 
 + \epsilon [ \partial_c\left(G_{\mu\nu} \partial_a X^\mu \partial_b X^\nu\right)\xi^c 
 \nonumber \\
 && +  G_{\mu\nu} \partial_c X^\mu \partial_b X^\nu \partial_a \xi^c
 + G_{\mu\nu}\partial_a X^\mu \partial_c X^\nu \partial_b \xi^c\, ]\, \} dx^a dx^b \ ,  
\eea
where we dropped the arguments for the sake of simplicity. Now it is easy to prove that
the latter expression can be written as
\be
h' = (h_{ab} + \epsilon \mathsterling_\xi h_{ab} ) dx^a dx^b \ ,
\ee
showing that under an infinitesimal diffeomorphism 
the components of $h_{ab}$ transform as
$h_{ab} \to h'_{ab} = h_{ab} + \mathsterling_\xi h_{ab}$. The finite version of
this infinitesimal diffeomorphism leads to the standard transformation law
of a $(0,2)$ rank tensor. Consequently, $h$ is a well-defined metric structure, 
induced by the map $X$ on the base space $M$. 
This proves the invariance of the action (\ref{genact})
under reparametrizations of the base space. It is worth noting that this invariance
is a consequence of the diffeomorphism invariance at the level of the 
Einstein-Hilbert action and the corresponding field equations. Indeed, for 
stationary axisymmetric fields in the Weyl-Lewis-Papapetrou representation,
diffeomorphism invariance of spacetime reduces to invariance with respect to 
arbitrary transformations relating the coordinates $\rho$ and $z$, and this is
exactly the reparametrization invariance of the base space as discussed above.

Finally, an important symmetry exists if the base space is 2-dimensional. 
In fact, in this case the change of the metric $\gamma^{ab}$ under an 
infinitesimal transformation (\ref{lie}) can be used in order to bring
it into the conformally flat form  $\gamma^{ab} = e^{2\phi(x)} 
\eta^{ab}$, where $\eta^{ab}$ ($a,b=1,2)$ is the (pseudo-)Euclidean metric,
and $\phi(x)$ is a smooth function. This property allows us to introduce
the Weyl transformation
\begin{equation}
\gamma_{ab} \to \gamma'_{ab} = e^{\sigma(x)}\gamma_{ab},
\end{equation}
which preserves the form of the generalized action (\ref{genact}).
In fact, the expression $\sqrt{|\gamma|}\gamma^{ab}$ is invariant under
a Weyl transformation since $\gamma'^{ab} = e^{-\sigma(x)}\gamma^{ab}$ 
and $\sqrt{|\gamma'|} =
e^{\sigma(x)}\sqrt{|\gamma|}$. This symmetry is associated with a local
rescaling of the metric $\gamma_{ab}$.

\subsection{Conservation laws}
\label{sec:claw}

As we mentioned before, the generalized action (\ref{genact}) includes an
interaction between the base space $N$ and the target space $M$, 
reflected on the fact that $G_{\mu\nu}$ depends explicitly on the coordinates 
of the base space. Clearly, this interaction must affect the conservation laws 
of the physical systems we attempt to describe by means of generalized
harmonic maps. 
To see this explicitly we calculate the covariant derivative 
of the generalized 
Lagrangian
density 
\be
{\cal L} = \sqrt{|\gamma|}\, \gamma^{ab}(x)\, \partial_a \, X^\mu
\partial_ b X^\nu  G_{\mu\nu}(X,x) \ ,
\label{genlag}
\ee
and replace in the result the corresponding
motion equations (\ref{gengeo}). Then, the final result can be written as
\be
\nabla_b  \widetilde T _a^{\ b} = - \frac{\partial {\cal L}}{\partial x^a} 
\label{claw1}
\ee
where $\widetilde T _a^{\ b}$ represents the canonical energy-momentum tensor
\be
\widetilde T_a^{\ b} = \frac{ \partial {\cal L} }{\partial (\partial_b X^\mu) }(\partial_a X^\mu) 
- \delta_a^b {\cal L} = 2 \sqrt{\gamma} G_{\mu\nu} \left( 
\gamma^{bc} \partial_a X^\mu \, \partial_c X^\nu - 
\frac{1}{2}\delta_a^b \gamma^{cd} \partial_c X^\mu \, \partial_d X^\nu\right).  
\label{emt}
\ee 
The standard conservation law is recovered only when the Lagrangian does not
depend explicitly on the coordinates of the base space. Even if we choose a 
flat base space $\gamma_{ab} = \eta_{ab}$, the explicit dependence of the metric
of the target space $G_{\mu\nu}(X,x)$ on $x$ generates a term that violates the
standard conservation law. This term is due to the interaction between 
the base space and the target space which, consequently, 
is one of the main characteristics of
the generalized harmonic maps introduced in this work. 

An alternative and more general 
definition of the energy-momentum tensor is by means of the variation 
of the Lagrangian density with respect to the metric of the base space, i.e.
\be 
T _{ab} = \frac{\delta {\cal L}}{\delta \gamma^{ab}} \ .
\ee
A straightforward computation shows that for the action under consideration here 
we have that $\widetilde T_{ab}= 2 T _{ab}$ so that the generalized conservation law 
(\ref{claw1}) can be written as 
\be
\nabla_b  T _a^{\ b} + \frac{1}{2} \frac{\partial {\cal L}}{\partial x^a} = 0 \ . 
\label{claw2}   
\ee
For a given metric on the base space, this represents in general a system of $m$ differential equations for the ``fields" $X^\mu$ which must be satisfied ``on-shell".

If the base space is 2-dimensional, we can use a reparametrization of $x$ to choose
a conformally flat metric, and the invariance of the Lagrangian density  under 
arbitrary Weyl transformations
to show that the energy-momentum tensor is traceless, $T _a^{\ a} =0$.



\begin{thebibliography}{99}
\bibitem{ernst} F. J. Ernst, {\it New formulation of the axially symmetric 
gravitational field problem}, {\it Phys. Rev.} {\bf 167} (1968) 1175;
F. J. Ernst, {\it New Formulation of the axially symmetric gravitational field problem II}
Phys. Rev. {\bf 168} (1968) 1415.

\bibitem{solutions} H. Stephani, D. Kramer, M. MacCallum, C. Hoenselaers, and E. Herlt, 
Exact solutions of Einstein's field equations, Cambridge University Press, Cambridge UK, 2003.

\bibitem{ours} H. Quevedo and B. Mashhoon, {\it Exterior gravitational field
of a rotating deformed mass}, {\it Phys. Lett. A} {\bf 109} (1985) 13;
H. Quevedo, {\it Class of stationary axisymmetric solutions 
of Einstein's equations in empty space}, {\it Phys. Rev. D} {\bf 33} (1986) 324;
H. Quevedo and B. Mashhoon, {\it Exterior  gravitational field of a charged
rotating mass with arbitrary quadrupole moment}, {\it Phys. Lett. A} {\bf 148} (1990) 149;
H. Quevedo, Multipole Moments in General Relativity - Static
and Stationary Solutions-, {\it Fort. Phys.} {\bf 38} (1990) 733;
H.Quevedo and B. Mashhoon {\it Generalization of Kerr spacetime}, {\it Phys. Rev. D}
{\bf 43} (1991) 3902.

\bibitem{maison} D. Maison, {\it Are the stationary, axially symmetric Einstein equations completely integrable?}, {\it Phys. Rev. Lett.} {\bf 41} (1978) 521. 



\bibitem{misner} C. W. Misner, {\it Harmonic maps as models for physical theories}, 
{\it Phys. Rev. D} {\bf 18} (1978) 4510. 

\bibitem{nicolai} D. Korotkin and H. Nicolai, {\it Separation of variables and 
Hamiltonian formulation for the Ernst equation}, {\it Phys. Rev. Lett.}
 {\bf 74} (1995) 1272. 

\bibitem{cnq01} D. Nu\~nez, H. Quevedo and A. S\'anchez, {\it Einstein's equations as 
functional geodesics}, {\it Rev. Mex. Phys.} {\bf 44} (1998) 440;
 J. Cortez, D. Nu\~nez, and H. Quevedo, {\it Gravitational fields 
and nonlinear sigma models}, {\it Int. J. Theor. Phys.} {\bf 40} (2001) 251.

\bibitem{nishino1} H. Nishino, {\it 
Stationary axisymmetric black holes,
N = 2 superstring, and self–dual gauge or gravity fields}, {\it Phys. Lett. B} 
{\bf 359} (1995) 77. 

\bibitem{nishino2} H. Nishino, {\it 
Axisymmetric gravitational solutions
as possible classical backgrounds
around closed string mass distributions}, {\it Phys. Lett. B},
{\bf 540} (2002) 125. 

\bibitem{burinskii} A. Ya. Burinskii, {\it Some properties of Kerr solution 
to low-energy string theory}, {\it Phys. Rev. D}, {\bf 52} (1995) 5826.


\bibitem{weyl} H. Weyl, {\it Zur Gravitationstheorie}, {\it Ann. Physik (Leipzig)}
 {\bf 54} (1917) 117.
\bibitem{lewis} T. Lewis, {\it Some special solutions of the equations of axially 
symmetric gravitational fields}, {\it 
Proc. Roy. Soc. London} {\bf 136} (1932) 176.

\bibitem{pap} A. Papapetrou, {\it Eine rotationssymmetrische L\"osung in de Allgemeinen
Relativit\"atstheorie}, {\it Ann. Physik (Leipzig)} {\bf 12} (1953) 309.

\bibitem{pol} J. Polchinski, String Theory: An introduction to the 
bosonic string, Cambridge University Press, Cambridge, UK, 2001. 

\bibitem{patquev05} L. Pati\~no and H. Quevedo, {\it Topological quantization of
gravitational fields}, {J. Math. Phys.}
{\bf 46} (2005) 22502.

\bibitem{bkz} V. A. Belinsky, I. M. Khalatnikov and E. M. Lifshitz, 
{\it Oscillatory approach to a singular point in the relativistic cosmology}, {\it
Adv. Phys.} {\bf 19} (1970) 525.

\bibitem{smq} A. S\'anchez, A. Mac\'\i as, and H. Quevedo, {\it 
Generating Gowdy cosmological models}, {J. Math. Phys.} {\bf 45} (2004) 1849.


\bibitem{san}
H. J. de Vega and N. G. Sanchez, {\it 
A new approach to string quantization in curved space-times}, {\it 
Phys. Lett. B} {\bf 197} (1987) 320;  
H.J. de Vega, I. Giannakis, and A. Nicolaidis, {\it 
String quantization in curved space-times: Null string approach}, {\it
Mod. Phys. Lett. A} {\bf 10} (1995) 2479; 
M. Maeno and S. Sawada, {\it String field theoru in curved space: A nonlinear sigma model
analysis}, {\it   Nucl. Phys. B} {\bf 306} (1988) 603; 
I. Bars, {\it Heterotic string models in curved space-time}, {\it
 Phys. Lett. B} {\bf 293} (1992) 315; 
N. G.  Sanchez, {\it  Advances in string theory in curved backgrounds: 
A synthesis report}, {\it
Int. J. Mod. Phys. A} {\bf 18} (2003) 2011.

\bibitem{gow} R. Gowdy,  {\it Gravitational waves in closed universes}, {\it
Phys. Rev. Lett.} {\bf 27} (1971) 826; 
{\it Vacuum space-times with two parameter spacelike isometry groups and compact invariant hypersurfaces: Topologies and boundary conditions}, {\it Ann.
Phys. (N.Y.)} {\bf 83} (1974) 203.

\end{thebibliography}
\end{document}